\documentclass[%
superscriptaddress,
twocolumn,
 amsmath,amssymb,
 aps,
prb,
floatfix,
]{revtex4-2}

\usepackage{graphicx}% Include figure files
\usepackage{dcolumn}% Align table columns on decimal point
\usepackage{bm}% bold math
\usepackage{xcolor}
\usepackage{multirow}
\usepackage{array}
\usepackage{ulem}
\usepackage{upgreek}

\definecolor{revised}{RGB}{0, 0, 0}
\definecolor{AMK}{RGB}{250, 0, 0}

\definecolor{Anna}{RGB}{55, 86, 255} %{55, 126, 184}

\newcommand{\FBO}{Fe$_3$BO$_6$}
\newcommand{\TN}{$T_\mathrm{N}$}
\newcommand{\Tsr}{$T_\mathrm{SR}$}
\newcommand{\Gtwo}{$\Gamma_2$}
\newcommand{\Gfour}{$\Gamma_4$}
\newcommand{\TO}{$T_0$}

\newcommand{\TT}{$\uptau_\mathrm{L}$}
\newcommand{\tsr}{$t_\mathrm{SR}$}

\begin{document}

\preprint{APS/123-QED}

\title{
Switching of an antiferromagnet\\ controlled by spin canting in a laser-induced hidden phase
}
\author{A.~V. Kuzikova}
\email{anna.kuzikova@mail.ioffe.ru}
\affiliation{%
 Ioffe Institute, 194021 St.~Petersburg, Russia}
%\author{?}
\author{N.~A. Liubachko}
\affiliation{%
Scientific-Practical Materials Research Centre, NAS of Belarus, 
220072 Minsk, Belarus}
\author{S.~N. Barilo}
\affiliation{%
Scientific-Practical Materials Research Centre, NAS of Belarus, 
220072 Minsk, Belarus}
\author{A.~V. Sadovnikov}
\affiliation{%
Laboratory ”Magnetic Metamaterials”, Saratov State University, 41001 Saratov, Russia}
\author{R.~V. Pisarev}
\affiliation{%
 Ioffe Institute, 194021 St.~Petersburg, Russia}
\author{A.~M. Kalashnikova}
\affiliation{%
 Ioffe Institute, 194021 St.~Petersburg, Russia}

\date{\today}

\begin{abstract}

During laser-induced phase transitions, fast transformations of electronic, atomic, and spin configurations often involve the emergence of hidden and metastable phases.
Being inaccessible under any other stimuli, such phases are indispensable for unveiling mechanisms and controlling the transitions.
We experimentally explore spin kinetics during the ultrafast first-order 90$^{\circ}$  spin-reorientation (SR) transition in a canted antiferromagnet \FBO, and reveal that the transition is controlled by canting between the magnetic sublattices. 
Laser-induced perturbation of the Dzyaloshinskii-Moriya interaction results in a change of the intersublattice canting within the first picoseconds, bringing \FBO{} to a hidden phase. 
Once this phase emerges, laser-induced heating activates precessional 90$^\circ$ spin switching.
The combination of spin canting and heating controls the final spin configuration comprising coexisting initial and switched phases.
The extended phase coexistence range is in striking contrast to the narrow SR transition in \FBO{} induced by conventional heating.

\end{abstract}

\maketitle

Understanding kinetics of phase transitions is foundational in many fields of physics and chemistry ranging from a mature topic of crystal growth to emergent topics of energy harvesting, smart materials, neuromorphic computing, etc.
The advancement of femtosecond laser technology noticeably augmented the possibilities to explore the physics of phase transitions \cite{delaTorre-RMP2021,Koshihara-PhysRep2022}.
Ultrafast optical excitation enables one to trace the intricate kinetics of the transition between the two equilibrium phases.
For first-order phase transitions, this kinetics may involve hidden and metastable phases that persist at time-scales ranging from subpicoseconds to milliseconds \cite{kiryukhin1997x, ichikawa2011transient, sun2020transient,  gao2022snapshots, ilyas2024terahertz, radu2011transient, white2025superheating}.
Access to the phases unattainable under equilibrium conditions renders possible differentiation between otherwise intertwined microscopical processes during the transition \cite{de2013speed,Wegkamp-ProgSS2015,li2022ultrafast,johnson-NPhys2024, lv2022unconventional}.
In turn, this insight facilitates the exploration of unconventional approaches to control phase transitions with various stimuli for ultrafast switching between states \cite{de2012coherent,Kurihara-PRL2018,Vaswani-PRX2020,afanasiev2021ultrafast,Truc-PNAS2024,Khokhlov-APLMater2024, maklar2023coherent}.
Therefore, the salient issue in the physics of laser-driven phase transitions is unveiling the transient states emerging after the excitation and establishing links between these states and properties of the induced phase.  

In magnetic media, spin-reorientation (SR) phase transitions involve 
pronounced change in the equilibrium direction of spins \cite{belov1976spin}. 
Naturally, laser-induced SR transitions \cite{kimel2002ultrafast,shelukhin2022spin,zhang2023optical,kuzikova2023laser} are seen as key contenders in the pursuit of all-optical ultrafast magnetic data processing \cite{kimel2004laser,schlauderer2019temporal,weiss2023discovery,Fichera2025PRX}.
Switching of antiferromagnets gains a particular attention as a base for perspective memory and logic technologies operating at the fundamental speed limits of magnetism \cite{rimmler2025non, han2023coherent,jungwirth2016antiferromagnetic}.
As multisublatice magnets, antiferromagnets may demonstrate SR transitions due to competing exchange, spin-orbital, and magnetic-dipolar energies 
with corresponding laser-driven spin kinetics  
spanning a wide range of characteristic timescales.
Therefore, comprehension of the first-order SR transition kinetics and identifying possible hidden phases paves a way towards controllable ultrafast spin switching.

In this Letter, we report on the laser-driven kinetics during the first-order SR transition in a canted antiferromagnet \FBO{} at non-cryogenic temperatures.
The SR transition in \FBO{} evolves 90$^\circ$ 
switching of a N\'eel vector and a magnetization, and a concomitant change of the magnetization value within a narrow temperature range of 0.5~K, posing yet unresolved question on a leading parameter of this phase transition.
We show that a decisive role in driving the SR transition is played by a decrease of the magnetization achieved through the laser-induced change of Dzyaloshinskii-Moriya coupling, resulting in a hidden phase. 
The magnetization in the hidden phase hinders or enables the complete 90$^\circ$ switching facilitated by laser-induced heating, and controls the latency of this process.

\begin{figure}
\includegraphics [width=0.47\textwidth] {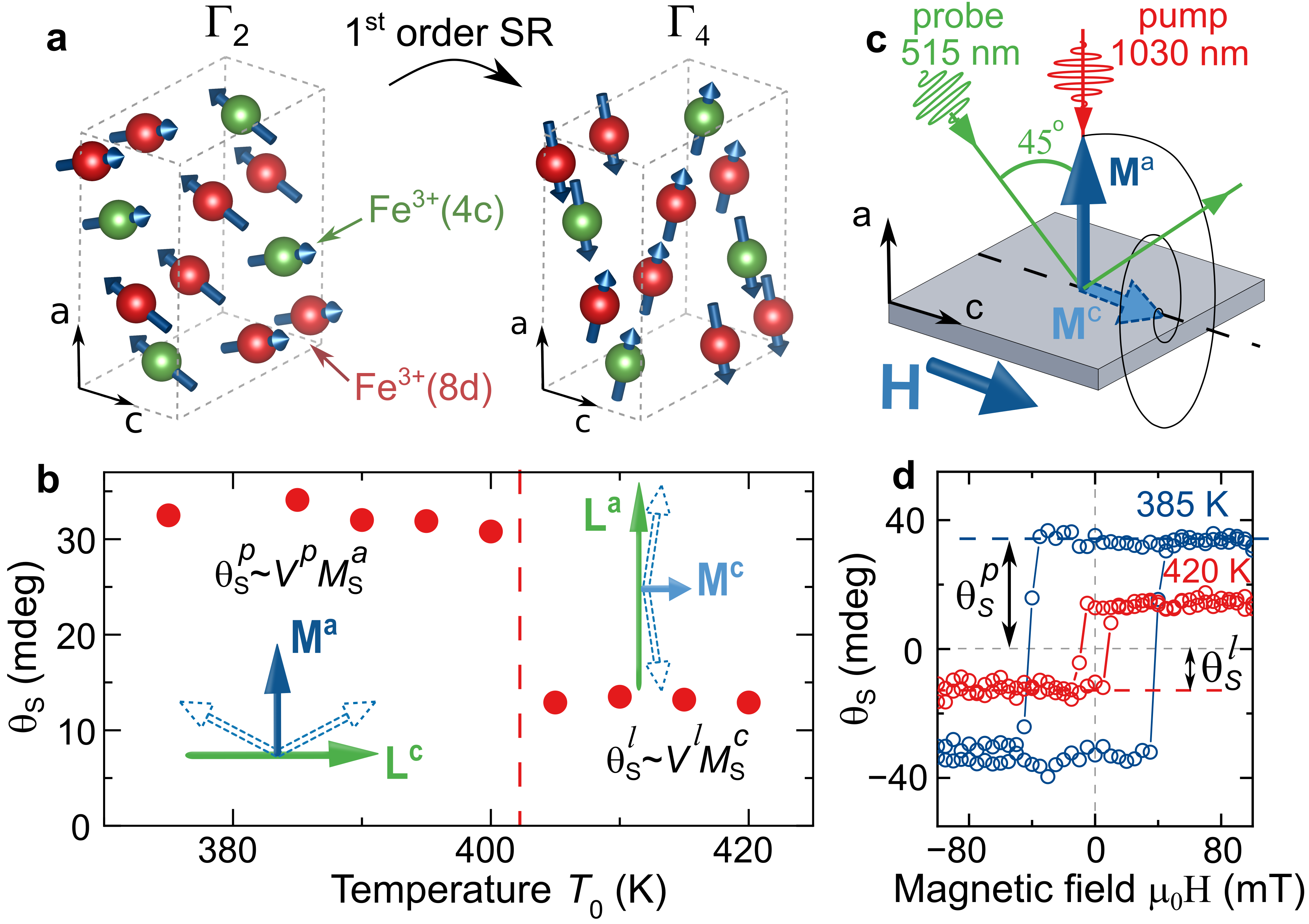}
\caption{\label{fig:structure} 
\textbf{Equilibrium SR transition in \FBO.}
(a) Crystal structure of \FBO, and spins arrangement at T <~\Tsr{} and T >~\Tsr. 
(b) Temperature dependence of MOKE $\uptheta^p_S$ and $\uptheta^l_S$ proportional to $M^a_S$ and $M^c_S$ below and above \Tsr, respectively.
Switching of the magnetization by 90$^\circ$ and the change of the intersublattice angle at SR transition are shown in insets.
(c) Experimental geometry for magneto-optical detection of the SR transition.
(d) Equilibrium magneto-optical hysteresis loops at temperatures below and above \Tsr.}
\end{figure}

Iron borate \FBO{} belongs to an orthorhombic space group $Pnma$ in which twelve Fe$^{3+}$ ions in the unit cell occupy two non-equivalent $8d$ and $4c$ octahedral positions [Fig.\ref{fig:structure}(a)] \cite{diehl1975refinement}. 
Magnetic moments of the iron ions within both positions are coupled antiferromagnetically and experience canting due to the Dzyaloshinskii-Moriya interaction (DMI). 
This results in a non-zero magnetization $M$. Below the SR transition temperature \Tsr~=~415~K~\cite{voigt1975temperature}, the phase is \Gtwo{} with a magnetization $M_S^a$ along the $a$-axis  [Fig.~\ref{fig:structure}(a, b)]. 
In a narrow temperature range of 0.5$-$3~K 
around \Tsr{} \cite{kamzin1993mossbauer,tsymbal2006orientation}, the first-order \Gtwo~$\rightarrow$~\Gfour{} transition occurs, and the magnetization switches by 90$^\circ$ from the $a$- to the $c$-axis.
Concomitantly, the magnetization $M_S^c$ is strongly reduced, $M_S^c/M_S^a = 0.6$ \cite{tsymbal2006orientation},
as the DMI parameters are different in the two phases \cite{mullerwiebus1977magnetic}.
The equilibrium transition is driven by temperature-induced variations of the competing DMI parameters rather than magnetocrystalline anisotropies \cite{mullerwiebus1977magnetic}.
\Gfour{} phase persists up to the N\'eel temperature \TN~= 508~K.

In the experiment, we used a 1.5-mm-thick plane parallel sample cut from a bulk Fe$_3$BO$_6$ single crystal grown from the flux \cite{prosnikov2022high}. 
The natural surface of the plate was perpendicular to the $a$-axis.
Upon SR transition, the magnetization switches between out-of-plane and in-plane alignments [Fig.~\ref{fig:structure}(c)].
We employ magneto-optical magnetometry based on a combination of polar (pMOKE) and longitudinal (lMOKE) Kerr effects measured with 170-fs 515-nm probe pulses at the angle of incidence of 45$^\circ$.
Fig.~\ref{fig:structure}(d) shows equilibrium MOKE hysteresis loops $\uptheta_S(\upmu_0H)$ below and above \Tsr{} measured with the magnetic field applied close to the $c$-axis [Fig.\ref{fig:structure}(c)]. 
The shape of the loop at $T_0=$ 385~K indicates that the strength of the in-plane field is insufficient to deviate the magnetization from the $a$-axis in the \Gtwo{} phase (Sec. I in Suppl. Mater.\cite{SM}).
Thus, at $T_0<~$\Tsr, $\uptheta_S$ originates from pMOKE and is proportional to the out-of-plane magnetization, $\uptheta_S^p=V^pM_S^a$.
Above \Tsr, $\uptheta_S$ stems from lMOKE and is proportional to the in-plane magnetization, $\uptheta_S^l=V^lM_S^c$.
In the experiment, the equilibrium SR transition is at \Tsr~=~403~K, where the abrupt change of $\uptheta_S$ (Fig.~\ref{fig:structure}(b) and Sec.~II in Suppl. Mater. \cite{SM}) is the result of the difference in the saturation magnetization in the \Gtwo{} and \Gfour{} phases. 
Additionally, it arises from the difference in the longitudinal $V^l$ and polar $V^p$ magneto-optical coefficients,
$V^l/V^p=0.64$.

\begin{figure}
\includegraphics [width=0.47\textwidth] {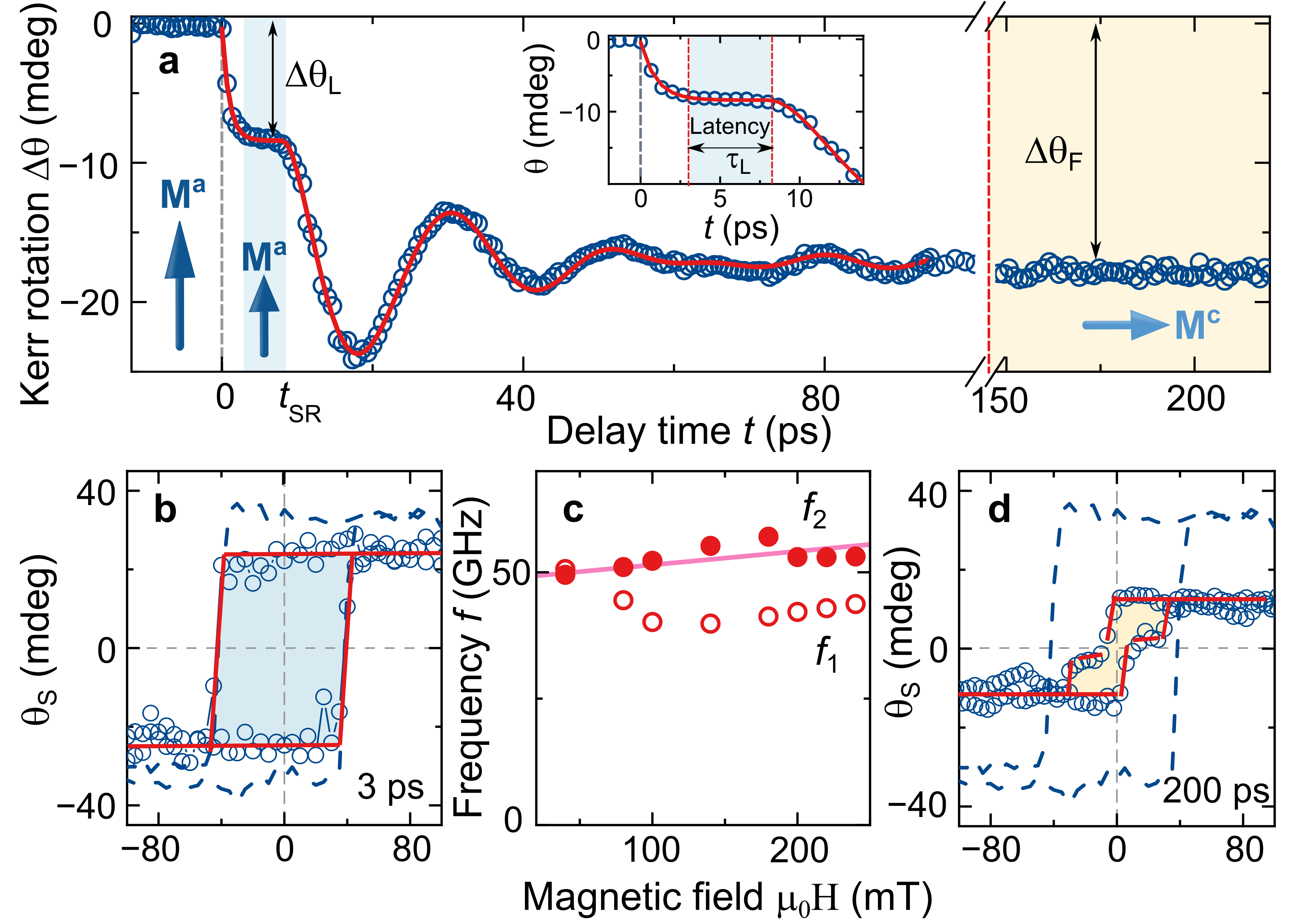}
\caption{\label{fig:Kerr_vs_time} 
\textbf{Kinetics of the laser-driven SR transition.}
(a)~Pump-probe signal $\Delta\uptheta(t)$ measured at $\upmu_0H~=~$200~mT, \TO~=~385~K, and $F$~=~0.5 J$\cdot$cm$^{-2}$ (symbols) and its fit (line). 
Inset zooms the time range of latency (see text for details).
(b, d)~MOKE hysteresis loops at the time delays (b)~$t=3$~ps and (d)~200~ps after laser excitation (symbols), and at $t=-10$~ps (dashed line). 
Solid lines are the guides to an eye.
(c) Field dependences of the precession frequencies $f_1$ (open symbols) and $f_2$ (closed symbols) obtained from $\Delta\uptheta(t,F)$ \TO~=~385 K, and of the qFMR frequency (line) calculated at \TO~=~430~K.}
\end{figure} 

Upon laser excitation of \FBO{} at initial temperature $T_0<~$\Tsr{}, a change of MOKE from $\uptheta_S^p$ to $\uptheta_S^l$ and the decrease of effective coercivity would signify a laser-induced SR transition.
We employed a femtosecond two-color magneto-optical pump-probe technique (Sec.~III in Suppl.~Mater. \cite{SM}) to monitor laser-induced change $\Updelta\uptheta$ of MOKE in the same geometry [Fig.~\ref{fig:structure}(c)]. 
170-fs 1030-nm pump pulses with a fluence $F$ were incident along the sample normal.
$\Updelta\uptheta$ was measured as a function of the pump-probe time delay $t$. 
The field-dependent signal was evaluated as $\Updelta\uptheta(t) = 0.5\left[\Updelta\uptheta(H_+; t) - \Updelta\uptheta(H_{-}; t)\right]$.
Fig.~\ref{fig:Kerr_vs_time}(a) shows $\Delta\uptheta$($t$) measured at $T_0=385$~K and pump pulse fluence $F$~=~0.5~J$\cdot$cm$^{-2}$.
The signal $\Updelta\uptheta(t)$ demonstrates 
a rapid exponential decrease by $\Updelta\uptheta_\mathrm{L}$ within $\approx$~3~ps after excitation followed by a latency \TT{} during which the signal remains constant [inset in Fig.~\ref{fig:Kerr_vs_time}(a)].
Following the latency, pronounced decaying oscillations start at \tsr{} and contain two frequencies, $f_1$ and $f_2$.
At $t >100$~ps, the signal settles at a final level $\Updelta\uptheta_\mathrm{F}$ and remains constant up to several nanoseconds (Sec.~IV in Suppl. Mater.~\cite{SM}).
The temporal evolution of $\Updelta\uptheta$ suggests that the precessional switching of the magnetization occurs, as expected at the laser-driven SR transition \cite{kimel2004laser}. 
However, it is preceded by the latency. 

We unveil the magnetic configurations 
preceding the precession and at the final state by measuring MOKE hysteresis loops at the time delays before, $t=-$10~ps, and after the pump pulse, $t=$~3 and 200 ps, at fluence $F=0.5$~J$\cdot$cm$^{-2}$ and temperature $T_0=385$~K [Fig.~\ref{fig:Kerr_vs_time}~(b, d)].  
At a negative time delay, the hysteresis loop is the same as the one measured under equilibrium conditions [Fig.~\ref{fig:structure}~(d)] at the same $T_0$, confirming that the system relaxes back to its initial state between two subsequent pump pulses.
During the latency, the coercivity is unchanged, showing that the magnetization remains aligned along the $a$-axis (Fig.~\ref{fig:Kerr_vs_time}(b) and Sec.~I in Suppl. Mater.~\cite{SM}). 
However, pMOKE at saturation is reduced by $\Updelta\uptheta_\mathrm{L}$. 
At the final state, $t = $~200 ps, the signal appears to be a superposition of the two types of hysteresis loops, characteristic of the \Gtwo{} and \Gfour{} phases [Fig.~\ref{fig:Kerr_vs_time}(d)].
This reveals that the SR transition indeed occurs at the given pump fluence, but with only a fraction $N^\mathrm{SW}$ of the material being switched to the \Gfour{} phase.
Thus, the signal at $t = $~200~ps is the sum of the pMOKE from the unswitched fraction, $(1-N^\mathrm{SW})(\uptheta_S^p$+$\Updelta\uptheta_\mathrm{L})$, and of the lMOKE from the switched fraction, $N^\mathrm{SW}\uptheta_S^l$. 

\begin{figure}
\includegraphics [width=0.48\textwidth] {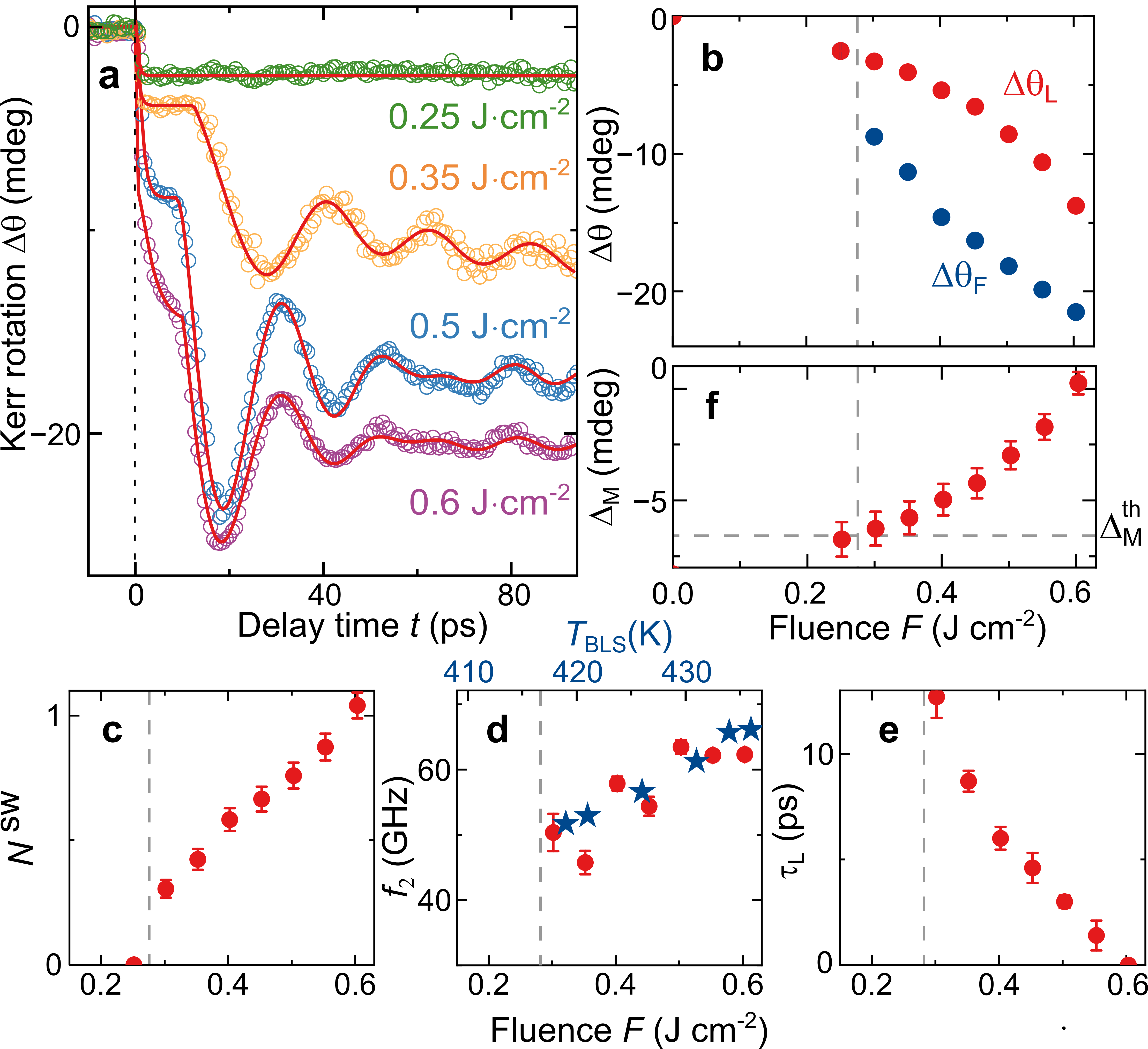}
\caption{\label{fig:precession}  
\textbf{SR transition at various laser pulse fluences.}
(a)~Pump-probe signals $\Delta\uptheta(t)$ (symbols) at $\upmu_0H~=~$200~mT, \TO~=~385~K at various pump fluences, and their fits (lines). 
(b)~Fluence dependences of $\Updelta\uptheta_\mathrm{L}$ during the latency \TT{} 
(red symbols) and  $\Updelta\uptheta_\mathrm{F}$ at $t>100$~ps (blue symbols).
Fluence dependences of (c)~the fraction $N^\mathrm{SW}$ of the material switched to the \Gfour{} phase,
(d)~the precession frequency $f_2$ (circles) collated with the temperature dependence of the qFMR frequency obtained from BLS (stars), 
(e)~the latency $\uptau_\mathrm{L}$, and (f)~the magnetization mismatch $\Updelta_\mathrm{M}$  [Eq.~\eqref{eq:mismatch}].
}
\end{figure}

To show that the complete SR transition with $N^\mathrm{SW}=1$ can be achieved, we examined $\Updelta\uptheta(t)$ traces measured at different $F$ [Fig.~\ref{fig:precession}(a)] and the dependences $\Delta\uptheta_\mathrm{L}(F)$ and $\Updelta\uptheta_\mathrm{F}(F)$ [Fig.~\ref{fig:precession}(b)].
We then calculated how the switched fraction $N^\mathrm{SW}$ changes with fluence using the relation 
$\Updelta\uptheta_\mathrm{F}= N^\mathrm{SW}(\uptheta^{l}_{s} - \uptheta^{p}_{s} -\Updelta\uptheta_\mathrm{L})+\Updelta\uptheta_\mathrm{L}$ (Sec. V in Suppl. Mater.~\cite{SM}).
Here we employed the fact that $\Updelta\uptheta_\mathrm{F}$ is the difference between the saturated signal at $t >$ 100ps and the equilibrium value $\uptheta_S^p$.
As seen in Fig.~\ref{fig:precession}(c) there is a threshold fluence $F_\mathrm{th}\approx 0.25$~J$\cdot$cm$^{-2}$ at which $N^\mathrm{SW}$ starts to increase from zero, that is characteristic for laser-induced phase transitions. 
Importantly, only at the highest fluence $F=0.6$~J cm$^{-2}$, the SR transition is completed, $N^\mathrm{SW}=1$, and only the \Gfour{} phase is present.

The final state at $t>100$~ps is achieved via the magnetization precession [Fig.~\ref{fig:Kerr_vs_time}(a)].
The corresponding oscillatory signal comprises two contributions. 
The first one with a higher amplitude and a lower frequency $f_1$ dominates at earlier time delays.
The second contribution possesses a lower amplitude and a higher frequency $f_2$.  
The field dependence of $f_2$ [Fig.~\ref{fig:Kerr_vs_time}~(c)] agrees well with the calculated values of a quasiferromagnetic resonance (qFMR) frequency in the \Gfour{} phase at 430~K (Sec.~VI in Supp. Mater. \cite{SM}).
This confirms that the fraction $N^\mathrm{SW}$ is switched to the \Gfour{} phase via precession.  
We attribute the $f_1$ oscillations to the early stage of the precession. 
It is triggered by the laser-induced 90$^\circ$ reorientation of the effective anisotropy axis and, thus, possesses a large amplitude beyond a linear regime of qFMR.

The precession frequency during the SR transition can be used to characterize the laser-induced phase \cite{afanasiev2021ultrafast,shelukhin2022spin}.
In Fig.~\ref{fig:precession}(d), we collate two dependences, namely the growth of $f_2$ with the pump fluence and the temperature dependence of the equilibrium qFMR frequency obtained from the Brillouin light scattering (BLS) (Sec.~VII in Suppl. Mater. \cite{SM}).
Evidently, the effective temperatures $T_\mathrm{F}=T_\mathrm{BLS}$ of the material reached after iteration at any pump pulse fluence $F\geq F_\mathrm{th}$ are above \Tsr.
Note that these effective temperatures are close to those estimated from the absorbed laser fluence (Sec.~ VIII in Suppl. Mater.~\cite{SM}).
On the one hand, this result confirms that the laser-driven SR transition takes place.
However, such laser-induced heating should be sufficient for the complete SR transition with $N^\mathrm{SW}=1$ even at $F_\mathrm{th}$ \cite{kamzin1993mossbauer,tsymbal2006orientation}, which is clearly not the case in our experiments.
This reveals that the laser-induced heating
is not a sufficient criterion for the complete SR transition.
To comprehend these contradictory observations, we turn to the analysis of the laser-induced state preceding the precession and described by the initial drop $\Delta\uptheta_\mathrm{L}$ and the latency \TT{}, the latter being a key signature of the laser-driven first-order transitions \cite{li2022ultrafast,sefidkhani2024metastability, afanasiev2017femtosecond}.

First, we note that the early evolution of $\Updelta\uptheta(t)$ at $t<~$\tsr{} does not follow the transient field-independent reflectivity change (Sec.~IX in Suppl. Mater. \cite{SM}), confirming that the signal is of a magnetic origin.
We extracted \TT{} at each fluence as a time range where  $d(\Updelta\uptheta) /d t=d^2 (\Updelta\uptheta) /d t^2$ = 0~(Sec.~X in Suppl. Mater.~\cite{SM}).
Noticeably, \TT{} decreases with increase of fluence [Fig.~\ref{fig:precession}(e)].  
As discussed above, $\Updelta\uptheta_\mathrm{L}$ does not originate from a deviation of the magnetization from its initial orientation.
Instead, $\Updelta\uptheta_\mathrm{L}$ reveals a reduction of the magnetization, $\Updelta\uptheta_\mathrm{L}=V^p\Updelta M^a$.
In a canted antiferromagnet, the weak magnetization depends on the sublattice magnetization and the angle between them \cite{dzyaloshinsky1958thermodynamic}.
In dielectrics, a degree of laser-induced demagnetization is governed by the lattice heating and the spin temperature following it~\cite{kimel2002ultrafast}, that gives a change of sublattice magnetization by just 15\% at the highest fluence (Sec.XI in Suppl. Mater.~\cite{SM}), that is less than the experimental values.
The canting angle between the antiferromagnetic sublattices is determined by a ratio of the DMI to the exchange parameter with a minor correction for the anisotropy parameters (see End Matter A). The reduced magnetization indicates the laser-induced decrease (increase) of the DMI (exchange) parameter. 

To evaluate how a change of the magnetization can yield the observed kinetics,
we recall that \Gtwo{} and \Gfour{} phases at equilibrium possess different magnetizations, as the DMI parameters are not equal for the two orientations.
We introduce a magnetization mismatch parameter $\Updelta_\mathrm{M}$ characterizing a difference between the magnetization $M_S^c$ in the \Gfour{} phase and $[M^a_S+\Updelta M^a(F)]$ at $t<~$\tsr.
This difference is multiplied by $V^l$ that enables evaluation of $\Updelta_\mathrm{M}$ based on the
measured equilibrium and transient MOKE values. Thus, we obtain 
\begin{align}
    \Delta_\mathrm{M}(F)=\uptheta_S^l-\frac{V^l}{V^p}\left[\uptheta_S^p+\Delta\uptheta_\mathrm{L}(F)\right],\label{eq:mismatch}
\end{align}
and this dependence is plotted in Fig.~\ref{fig:precession}(f).
The magnetization mismatch $\Delta_\mathrm{M}^\mathrm{th}$ at $F_\mathrm{th}$ is found to be equal to the discontinuity of the equilibrium magneto-optical signal at \Tsr{} [Fig.~\ref{fig:structure}(b)] expressed according to Eq.\eqref{eq:mismatch}.
At lower \TO, the threshold fluence $F_\mathrm{th}$ increases, while the corresponding $\Delta_\mathrm{M}^\mathrm{th}$ remains the same, supporting the suggested scenario (Sec. XII in Suppl. Mater.~\cite{SM}).
The completed laser-induced SR transition ($N^\mathrm{SW}=1$) is realized when $\Updelta_\mathrm{M}=0$, i.e. 
$[M^a_S+\Updelta M^a(F)]$ 
matches $M_S^c$.
Concomitantly, the latency \TT{} reduces to zero [Fig.~\ref{fig:precession}(e)].

\begin{figure}
\includegraphics [width=0.46\textwidth] {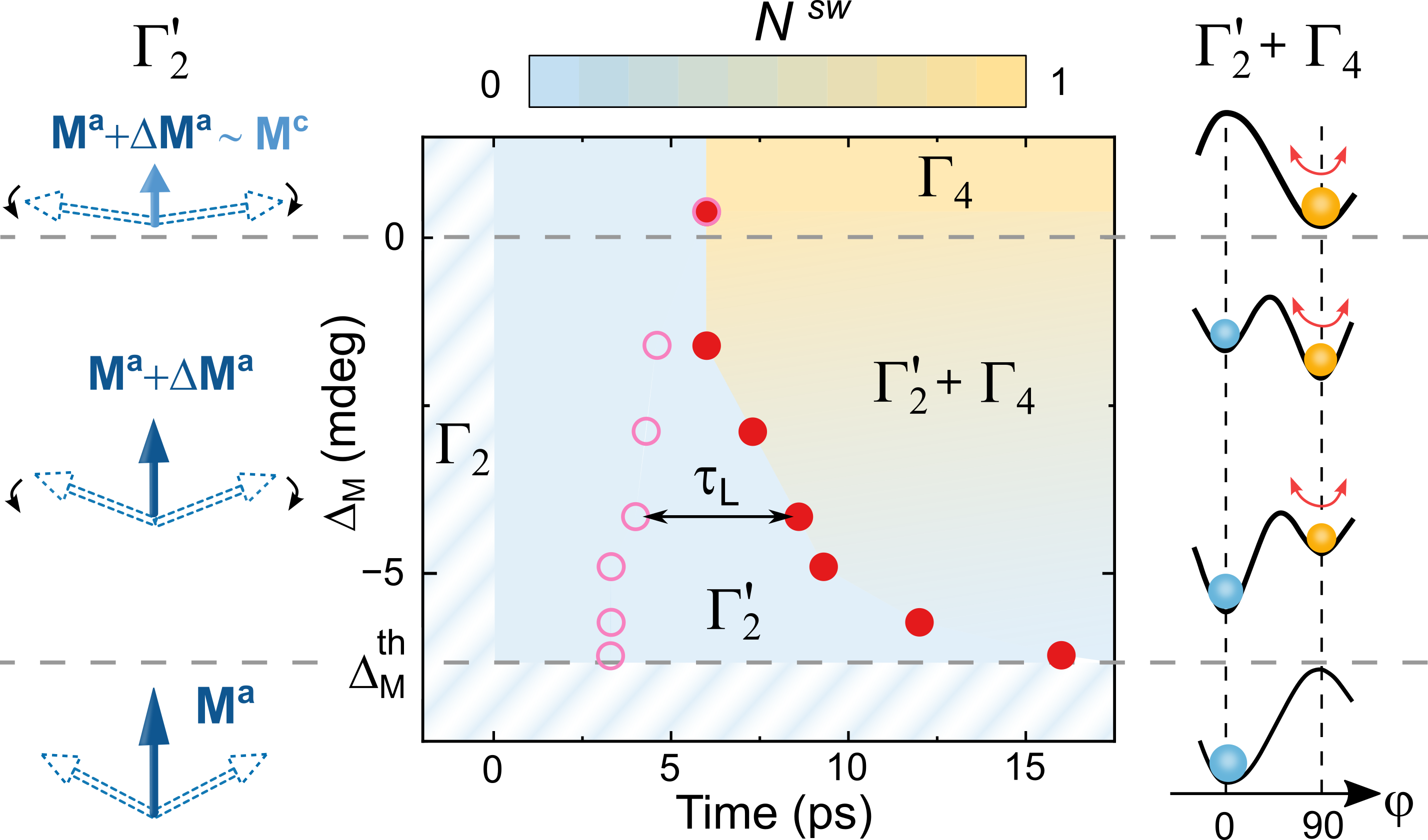}
\caption{\label{fig:latency} 
\textbf{Phase diagram of the laser-induced SR transition.} % in \FBO.}
Phase diagram~$t-\Delta_\mathrm{M}$ of the laser-induced SR transition based on the fluence dependence at $T_0=385$~K with the color-coded switched fraction $N^\mathrm{SW}$.
Open and closed symbols mark the start and the end (\tsr) of the latency.
Left-hand side illustrates the spins configuration at $t <$~\tsr.
Right-hand side shows the thermodynamics potential resulting from the DMI change, and the precession at $t >$~\tsr.
}
\end{figure}

To establish a relation between the magnetization mismatch, the extended range of the phases coexistence, and the latency, we use a phenomenological description of the SR transition \cite{belov1976spin}. 
The anisotropic part of the thermodynamic potential takes a form:
\begin{align}
    \Phi  =  K_1\cos^2{\upvarphi} + K_2 \cos^4{\upvarphi},\label{eq:energy}
\end{align}
where $K_1$ and $K_2$ are effective anisotropy parameters, $\upvarphi$ is an angle between the magnetization $M$ and the $a$-axis. % in the $ac$ plane.
In \FBO{}, the variation of $K_{1}$ and $K_{2}$ is predominantly governed by the changes of the DMI parameters and the magnetization (see End Matter A).
The established above link between the magnetization mismatch and SR transition allows us to conjecture that the reduced magnetization after the laser excitation primarily results from  the decrease of the DMI rather than the increase of the exchange parameter.
The SR transition is of the first order if $K_2<0$. 
At $K_1<0$ there is only the \Gtwo{} phase, while at $K_1+2K_2>0$ there is only the \Gfour{} phase.
At $K_1>0$ and $K_1+2K_2<0$ the two phases coexist.
In equilibrium a narrow range of the SR transition signifies that $K_1=0$ is satisfied at $M=M_S^a$($\approx$\Tsr)$~\propto d_a$, while $K_1+2K_2=0$ requires  $M=M_S^c$($\approx$\Tsr)$~\propto d_c$, where $d_a\neq d_c$ are the effective parameters of the DMI when the magnetization is aligned along the $a$- or $c$-axis, respectively (see Sec.~VI A in Suppl. Mater.~\cite{SM}).

Right after the laser excitation, the magnetization remains aligned along the $a$-axis, 
and the fulfillment of the two conditions on $K_1$ and $K_1+2K_2$ is controlled solely by $M^a$.
This leads to a phase diagram in the $t-\Updelta_\mathrm{M}$ coordinates shown in Fig.~\ref{fig:latency}.
At $F=F_\mathrm{th}$, 
$\Delta_\mathrm{M}^\mathrm{th}$ signifies that $[M^a_S+\Delta M^a(F_\mathrm{th})]$ becomes equal to $M^a_S$(\Tsr).
This yields the phase boundary $K_1=0$ similarly to the equilibrium case.
At $F>F_\mathrm{th}$, 
$[M^a_S+\Delta M^a(F)]$ is different from either of the equilibrium magnetizations $M_S^a$ and $M_S^c$.
Therefore, the laser-induced phase during the latency is a hidden phase designated as \Gtwo$'$.
Under this condition, $K_1+2K_2<0$, 
and only a partial switching of the magnetization to the $c$-axis occurs after a delay.
This effectively extends the range of (\Gtwo$'$+\Gfour) phases coexistence, and we obtain $N^\mathrm{SW}<1$ in a wide range of laser fluences.
The condition for the complete SR transition, $K_1+2K_2=0$, is reached only if $[M^a+\Delta M^a(F)]=M^c_S$, i.e.
$\Delta_\mathrm{M}=0$.
This is achieved at $F$ = 0.6~J$\cdot$cm$^{-2}$, where the
transition \Gtwo~$\rightarrow$~\Gfour{} occurs
without latency.
Expansion of the (\Gtwo$'$+\Gfour) phase coexistence range is well reproduced in simulations assuming laser-induced change of $d_a$ as a leading mechanism (Sec.~VI~C in Suppl. Mater.~\cite{SM}).

The analysis demonstrates that the change of the DMI parameter $d_a$ by a laser pulse with the fluence above $F_\mathrm{th}$ prepares the state with the two-minima potential, corresponding to $K_1>0$ and $K_1+2K_2<0$ (Fig.~\ref{fig:latency}). 
For the switching of the magnetization from the $a$-axis ($\upvarphi=0$) to the $c$-axis ($\upvarphi=90^\circ$), the barrier between the minima should be overcome.
The height of the barrier $\Delta\Phi$ is controlled by $\Updelta_\mathrm{M}$.
Following the N\'eel-Brown model for a macrospin switching \cite{brown1963thermal}, the ratio between $\Delta\Phi$ and the thermal energy $k_\mathrm{B}T_\mathrm{F}$ defines the probability and time required for the switching between the two minima.
Then, the gradient of $N^\mathrm{SW}$ in the $\Delta_\mathrm{M} - t$ phase diagram should change when the range of the corresponding $T_\mathrm{F}$ is changed.
This is realized in an experiment performed at various initial temperatures and fixed fluence $F=0.5$~J$\cdot$cm$^{-1}$ (see End Matter B for details).
It is found that $\Delta\uptheta_L$ is defined by the fluence and only weakly depends on the $T_0$. $\Delta_M$ is weakly dependent on $T_0$, while $T_\mathrm{F}$ increases linearly with $T_0$.
Then $N^\mathrm{SW}$ spans from 0 to 1 in the narrower range of $\Delta_\mathrm{M}$ because of higher $T_\mathrm{F}$.

To address a possible mechanism of the laser-induced DMI change, we stress that the change of intersublattice angle ($\propto\Updelta\uptheta_\mathrm{L}$) is found to be a superlinear function of the laser fluence [Fig.~\ref{fig:precession}(b)], while showing a weak dependence on temperature pointing to a nonthermal mechanism of the DMI change (see End Matter B). The latter is further supported by the absence of a correlation between $\Updelta\uptheta_\mathrm{L}$ at a particular fluence [Fig.~\ref{fig:precession}(b)] and the change of the static Kerr rotation due to an equivalent heating [Fig.~\ref{fig:structure}(b)].
The time during which $\Updelta\uptheta$ reaches $\Updelta\uptheta_\mathrm{L}$ takes values between 3 and 6~ps, as plotted in the phase diagram (Fig.~\ref{fig:latency}, open symbols).
This increase correlates with the softening of the quasi-antiferromagnetic resonance (qAFMR) mode starting from $\approx$~300~GHz at \TO~=~385~K \cite{arutyunyan1989magnetic} that can be caused by the decrease of the DMI parameter $d_a$. 
Thus, it is plausible that the spin canting change proceeds via a single cycle of strongly damped qAFMR \cite{mikhaylovskiy2015ultrafast}.
We speculate that the decrease rate of $d_a$ is comparable or faster than this time.
Possible microscopical mechanism of the non-thermal DMI change in iron oxides is proposed in Ref.~\cite{Mikhaylovskiy-PRL2020} and relies on the modified overlap of the 3$d^5$ electron wavefunctions upon excitation of the $d-d$ transitions in Fe$^{3+}$ ions. 
In \FBO, these transitions are close to the spectral range of the pump pulses \cite{andlauer1977optical}.
At the used high fluences, an impact of such excitations can be pronounced and, moreover, is not hindered by the moderate heating owing to the large heat capacity of \FBO{} at the studied $T_0$ (Sec.~VII in Suppl. Mater. \cite{SM}).
Further, the \Gtwo{}$'$ phase is present at the timescales up to 15~ps, and the (\Gtwo{}$'$+\Gfour) persists upto several nanoseconds, showing that $d_a$ retains its reduced value within this time range.
This suggests that the hidden \Gtwo{}$'$ phase is stabilized at temperatures $T_\mathrm{F}$. 

In conclusion, we demonstrated laser-induced first-order SR transition \Gtwo~$\rightarrow$~\Gfour{} in the canted antiferromagnet \FBO, at which the magnetization is switched by 90$^{\circ}$ from the $a$- to $c$-axis and reduced by $\approx35$\%.
The SR transition is enabled by the laser-induced spin canting mainly originating from the DMI parameter change, which creates a hidden phase with magnetization neither of the \Gtwo~or~\Gfour{} equilibrium phases.
The spin canting in the hidden phase is controlled by the laser fluence and is suggested to be predominantly of nonthermal origin. 
The switching to the final state is facilitated by laser-induced heating.
Nevertheless, the hidden phase appears to be determinative for the spin kinetics during the entire transition from pico- to nanoseconds by significantly broadening the range of magnetic-symmetry-different phases coexistence. 
The latter is inherent for the first-order phase transitions and defines their functionality for e.g. conventional and neuromorphic computation~\cite{schofield2023harnessing, markovic2020physics}. 
Our finding brings to the fore the feasibility of tuning this range dynamically by means of a laser-driven hidden state.
As spin canting and heating have distinct effects on the SR transition, our results pave the way to extended control of spin switching in \FBO. 
Additional stimuli, such as optical and THz electric fields \cite{mentink2015ultrafast,Mikhaylovskiy-PRL2020,fedianin2023selection} or coherent phonons \cite{maehrlein2018dissecting,afanasiev2021ultrafast,Bossini_2023} can be used to control the spin canting by perturbing the exchange coupling along with DMI.
Our results pose a question on theoretical comprehension of the laser-driven change of the intrinsic DMI. On the other hand, it suggests that engineering SR transitions in heterostructures by means of interfacial DMI \cite{yang-PRL2015,caretta-NComm2020} can be exploited for ultrafast switching.

\section*{DATA AVAILABILITY}
The data generated in this study have been deposited in the Zenodo data base \cite{data_ref}.

\begin{acknowledgments}
We thank L.~A.~Shelukhin for the help with pump-probe experiments, M.~P.~Scheglov for the sample characterization, and M.~A.~ Prosnikov for insightful discussions.
The work of A.V.K. and A.M.K was supported by the Russian Science Foundation (grant No.~23-12-00251).
The work of S.N.B. was supported by the the Belarusian Republican Foundation for Fundamental Research (grant No.~25KI-070).
\end{acknowledgments}

\bibliography{apssamp}% Produces the bibliography via BibTeX.

\section*{End Matter}
\subsection{Phenomenological theory of the SR transition}

For a theoretical description of the SR transition between the two magnetic states of Fe$_3$BO$_6$, an expansion of the thermodynamic potential 
for the two sublattice antiferromagnet with the $Pnma$ group is used \cite{ozhogin1972statics}:
\begin{multline}
    \Upphi  =  \frac{E}{2} m^ 2 - d_a m_a l_c + d_c m_c l_a + \frac{a_1}{2} l^2 _c + \frac{c_1}{2} l^2 _a + \\
    \frac{a_2}{4} l^4 _c + \frac{c_2}{4} l^4 _a + \frac{f}{2} l^2 _c l^2 _a.
    \label{eq:DMI}
    \end{multline}
where $\textbf{m} = (\textbf{M}_1 + \textbf{M}_2)/M_0$, and $\textbf{l} = (\textbf{M}_1 - \textbf{M}_2)/M_0$, are the ferromagnetic and antiferromagnetic vectors, respectively.
$\textbf{M}_1$ and $\textbf{M}_2$ are the magnetic moments of the sublattices, $M_0$ is their magnitude at the given temperature.
$E/2$ is effective exchange parameter, $d_a\neq d_c$ are effective parameters of the DMI, when the magnetization is aligned along the $a$- (\Gtwo{} phase; $\upvarphi=0$) or $c$-axis (\Gfour{} phase; $\upvarphi=90^\circ$), respectively; $a_1$, $c_1$, $a_2$, $c_2$, $f$ are the bilinear and biquadratic magnetocrystalline anisotropy parameters.

To describe the connection between phase transition conditions and the DMI parameters, we use the formalism introduced in Eq.~\eqref{eq:energy}. 
Taking into account Eq.~\eqref{eq:DMI}, the dependence of the effective anisotropy constants $K_1$, $K_2$ in Eq.~\eqref{eq:energy} on the DMI parameters and the magnetization is obtained:
\begin{align}
    K_1  = d_c m - d_a m + a_2 m^2 - \frac{a_1}{2} + \frac{c_1}{2} - \frac{a_2}{2} + \frac{f}{2}; \nonumber  \\ 
    K_2  = - \frac{a_2}{2} m^2 - \frac{c_2}{2} m^2 + \frac{a_2}{4} + \frac{c_2}{4} - \frac{f}{2},
    \label{eq:K12}
\end{align}   
where conditions $l^2+m^2=$~1 and $m\ll$~1 were used.

At $K_1>0$ and $K_1+2K_2<0$ the two phases coexist.  
At equilibrium, the temperature range of the phase coexistence is below 3~K \cite{kamzin1993mossbauer,tsymbal2006orientation}.
From Eq.~\eqref{eq:K12} it follows that the condition $K_1=0$ is satisfied at $m=m_S^a$($\approx$\Tsr), while $K_1+2K_2=0$ requires  $m=m_S^c$($\approx$\Tsr) 
with
\begin{align}
 m =  \Bigl\{
\begin{array}{ll}
m^a=d_a(E-c_2)^{-1} & \textrm{at \Gtwo}\\
m^c=d_c(E-a_2)^{-1} & \textrm{at \Gfour}
\end{array} \Bigr..\label{eq:M}
\end{align}

\subsection{Role of the laser-induced heating}
\begin{figure}
\includegraphics [width=0.48\textwidth] {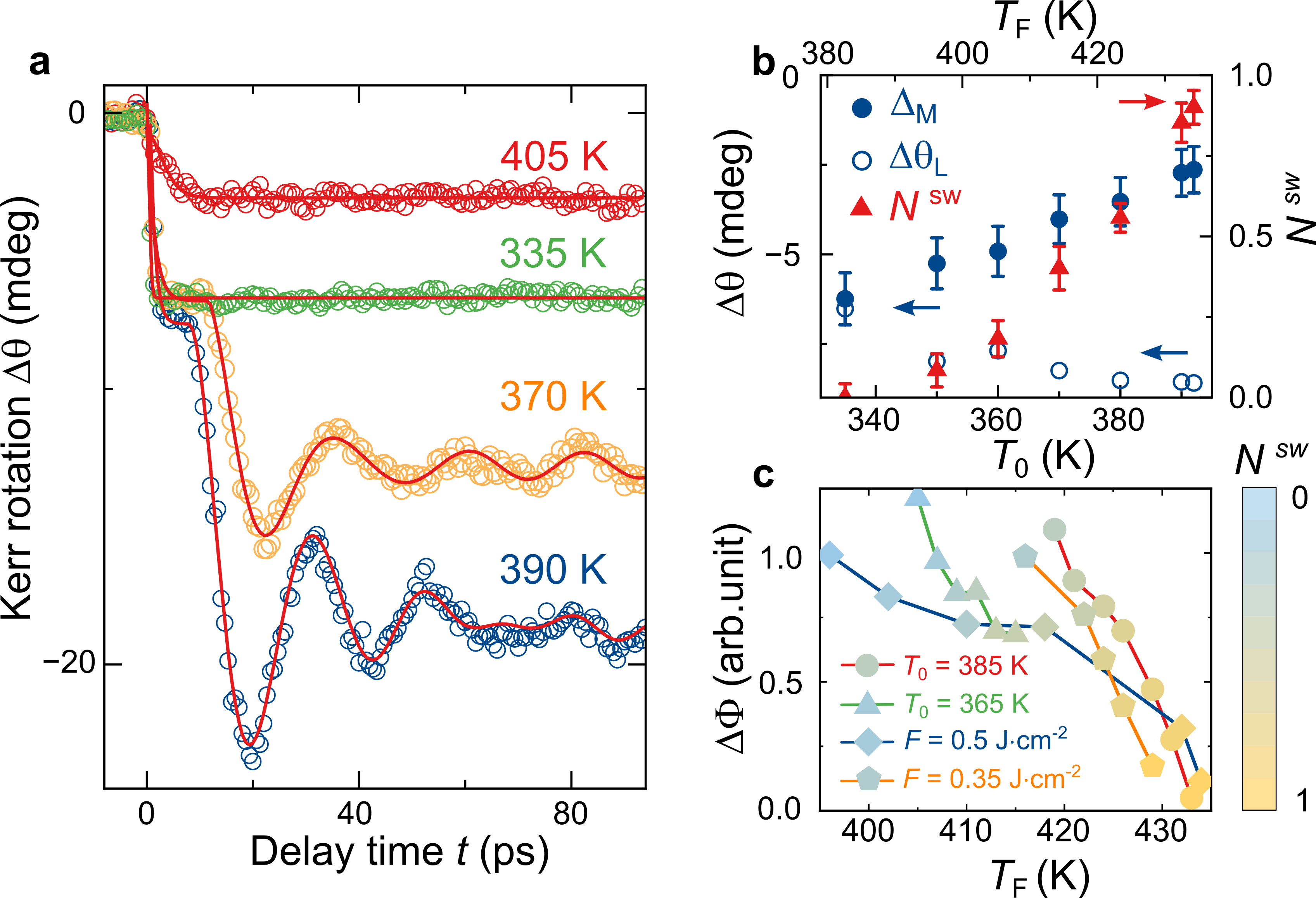}
\caption{\label{fig:end} 
%\textbf{SR transition at various initial temperatures.}
(a)~Pump-probe signals $\Delta\uptheta(t)$ (symbols) at $\upmu_0H~=~$200~mT, $F$ = 0.5~J$\cdot$cm$^{-2}$ at various initial temperature, and their fits (lines). 
(b)~The magnetization mismatch $\Delta_\mathrm{M}$ (closed circles), $\Delta\theta_\mathrm{L}$ (open circles), and the switched fraction $N^\mathrm{SW}$ (triangles) 
vs. initial and final temperatures as measured at the fixed fluence $F$ = 0.5~J$\cdot$cm$^{-2}$.
(c) 2D plot of $N^\mathrm{SW}$ vs. barrier $\Updelta\Upphi$ and the final temperature $T_\mathrm{F}$ as obtained from the fluence dependences at $T_0$~=~385 (circles) and 365~K (triangles) and the temperature dependeces at $F$~=~0.5 (diamonds) and 0.35~J$\cdot$cm$^{-2}$ (pentagons).}
\end{figure}

The role of laser-induced heating becomes evident from the dependences of $\Delta\uptheta_\mathrm{L}$, $\Delta_\mathrm{M}$ and $N^\mathrm{SW}$ on $T_0$ and on the corresponding $T_\mathrm{F}$ obtained from the experiments at fixed pump fluence $0.5$~J$\cdot$cm$^{-2}$ and various $T_0$ (Fig.~\ref{fig:end}(a) and Sec.~XII in Suppl. Mater. \cite{SM}).
$\Delta\uptheta_\mathrm{L}$ is found to be only weakly dependent on the initial $T_0$ [Fig.~\ref{fig:end}(b)] that is in contrast with its superlinear fluence dependence [Fig.~\ref{fig:precession}(b)]. This allows us to conclude that the initial change of magnetization $m$ and, therefore, of DMI, is predominantly non-thermal, since a thermally-driven process would demonstrate linear dependence on the fluence as well. Indeed, in a case of a purely thermal effect both fluence and temperature dependences would be linear, as evident from, e.g., a thermally-induced strain effect seen in reflectivity measurements (Sec.~IX in Suppl. Mater. \cite{SM}).

Correspondingly,
the magnetization mismatch $\Delta_\mathrm{M}$ is weakly altered upon increase of $T_0$ [Fig.~\ref{fig:end}(b)].
The final temperature $T_\mathrm{F}$, in turn, changes linearly with $T_0$ (Sec.~VII in Suppl. Mater.~\cite{SM}).
$N^\mathrm{SW}$ shows a pronounced increase with $T_0$ and 
$T_\mathrm{F}$ signifying that the switching that follows the initial non-thermal process, is thermally activated.
Based on this model, we evaluate the barrier height $\Delta\Phi$ from \TT{} and $T_\mathrm{F}$ as 
\begin{align}
\uptau_{\mathrm{L}} \propto \mathrm{exp} \left[ \frac{\Updelta\Upphi}{(k_{\mathrm{B}} T_{\mathrm{F}})}\right].
\end{align}

The sets of \TT{} and $T_\mathrm{F}$, along with $N^\mathrm{SW}$, were obtained from the fluence dependences measured at \TO~=~365 and 385~K and from the temperature dependences at $F$~=~0.35 and 0.5~J$\cdot$cm$^{-2}$ (Sec.~XII in Suppl. Mater. \cite{SM}).
2D plot in Fig.~\ref{fig:end}(c) shows how $N^\mathrm{SW}$ changes with $\Delta\Phi$ and $T_\mathrm{F}$.
All the data fall within the same trend confirming the roles played in the laser-induced SR transition by the change of the DMI and the heating.
%
% ****** End of file apssamp.tex ******

\end{document}